\documentclass{ws-book9x6}        

\usepackage{makeidx}
\makeindex

\begin{document}

%%%%%%%%%%%%%% Please do not remove these lines %%%%%%%%%%%%%%%%%%%%%
%\titlepages %pls. leave this as it is. 
%%%%%%%%%%%%%%%%%%%%%%%%%%%%%%%%%%%%%%%%%%%%%%%%%%%%%%%%%%%%%%%%%%%%%

\setcounter{page}{1}
\setcounter{chapter}{9}

%%Please change your runningheads for the {Book Title}{Chapter Title} 
%%at the \markboth command here % 

\chapter{Towards Understanding\\ the Origin of Genetic Languages}
%        Why do living organisms use 4 nucleotide bases and 20 amino acids?
\centerline{Apoorva D. Patel}
%        Centre for High Energy Physics and
%        Supercomputer Education and Research Centre
%        Indian Institute of Science
%        Bangalore 560012, India
%        Invited talk at the Indo-US Shared Vision Workshop on
%        Soft, Quantum and Nano Computing (SQUAN-2007),
%        Dayalabagh, Agra, 22-25 February 2007.

\markboth{Quantum Aspects of Life}{Towards Understanding the Origin of Genetic Languages}  %CHANGE HERE

\vspace{5mm}
% Sing a song of sixpence a pocket full of rye,
\rightline{\it ``\ldots four and twenty blackbirds baked in a pie \ldots"}
% When the pie was opened, the birds began to sing.
% Now, wasn't that a dainty dish to set before the king?
\vspace{5mm}

Molecular biology is a nanotechnology that works---it has worked
for billions of years and in an amazing variety of circumstances.
At its core is a system for acquiring, processing and communicating
information that is universal, from viruses and bacteria to human
beings. Advances in genetics and experience in designing computers
have taken us to a stage where we can understand the optimisation
principles at the root of this system, from the availability of
basic building blocks to the execution of tasks. The languages of
DNA and proteins are argued to be the optimal solutions to the
information processing tasks they carry out. The analysis also
suggests simpler predecessors to these languages, and provides
fascinating clues about their origin. Obviously, a comprehensive
unraveling of the puzzle of life would have a lot to say about
what we may design or convert ourselves into.

\section{The Meaning of It All}

I am going to write about some of the defining characteristics of life.
Philosophical issues always arise in discussions regarding life, and
I cannot avoid that. But let me state at the outset that such issues
are not the purpose of my presentation. I am going to look at life as
an exercise in information theory, and extend the analysis as far as
possible.

Let me begin with the textbook answer to the question made famous by
Schr\"odinger \cite{schrodinger}: {\it What is life?}
Life\index{life} is fundamentally a non-equilibrium process, commonly
charaterised in terms of two basic phenomena. One is ``metabolism".
Many biochemical processes are needed to sustain a living organism.
Running these processes requires a continuous supply of free energy,
which is extracted from the environment. (Typically this energy is in
electromagnetic or chemical form, but its ultimate source is gravity---the
only interaction in the universe that is not in equilibrium.) The other
is ``reproduction". A particular physical structure cannot survive forever,
because of continuous environmental disturbances and consequent damages.
So life perpetuates itself by a succession of generations.

It is obvious that both these phenomena are sustaining and protecting and
improving something, often against the odds. So let us figure out what is
it that is being sustained and protected and improved.

All living organisms are made up of atoms. These atoms are fantastically
indestructible. In all the biochemical processes, they just get rearranged
in different ways. Each of us would have a billion atoms that once belonged
to the Buddha, or Genghis Khan, or Isaac Newton---a sobering or exciting
realisation depending on one's frame of mind! We easily see that it is
not the atoms themselves but their arrangements in complex molecules,
which carry biochemical information\index{biochemical information}.
In the flow of biochemical processes, living organisms synthesise
and break up various molecules, by altering atomic arrangements.
The biochemical information resides in what molecules to use where,
when and how. Characterisation of this information is rather abstract,
but central to the understanding of life. To put it succinctly:
\begin{center}
Hardware is recycled, while software is refined!
\end{center}
At the physical level, atoms are shuffled, molecules keep on changing,
and life goes on. At the abstract level, it is the manipulation and
preservation of information that requires construction of complex
structures. Information\index{information theory} is not merely ``a"
property of life---it is ``the" basis of life.

Now information is routinely quantified as entropy of the possible forms
a message could take \cite{shannon}. What the living organisms require,
however, is not mere information but information with meaning. A random
arrangement of components (e.g. a gas) can have large information, but it
is not at all clear how that can be put to any use. The molecules of life
are destined to carrying out specific functions, and they have to last
long enough to execute their tasks. The meaning of biological information
is carried by the chemical properties of the molecules, and a reasonably
stable cellular environment helps in controlling the chemical reactions.
What the living organisms use is ``knowledge"\index{knowledge},
\begin{center}
Knowledge = Information + Interpretation.
\end{center}
Knowledge has to be communicated using a language\index{language}.
A language uses a set of building blocks (e.g. letters of an alphabet)
whose meaning is fixed, and whose variety of arrangements (invariably
aperiodic) compose different messages. It is the combination of
information and interpretation that makes languages useful in practice.

Thus to understand how living organisms function, we need to focus on the
corresponding languages whose interpretation remains fixed, while all
manipulations of information processing go on. A practical language is never
constructed arbitrarily---criteria of efficiency\index{language!efficiency}
are always involved. These criteria are necessarily linked to the tasks
to be implemented using the language, and fall into two broad categories.
One is the stability of the meaning, i.e. protection against error
causing fluctuations. And the other is the efficient use of physical
resources, i.e. avoidance of unnecessary waste of space, time, energy
etc. while conveying a message. The two often impose conflicting demands
on the language, and the question to investigate is: {\it Is there
an optimal language for a given task, and if so how can we find it?}
From the point of view of a computer designer, the question has two parts:
\begin{center}
Software: What are the tasks? What are the algorithms?\\
Hardware: How are the operations physically implemented?
\end{center}
It goes without saying that the efficiency of a language depends both
on the software and the hardware.

In the computational complexity analysis, space and time resources are
often traded off against each other, and algorithms are categorized as
polynomial or non-polynomial (usually exponential). In the biological
context, however, the efficiency considerations are not quite the same.
Time is highly precious, while space is fairly expendable. Biological
systems can sense small differences in population growth rates, and even
an advantage of a fraction of a percent is sufficient for one species to
overwhelm another over many generations. Spatial resources are frequently
wasted, that too on purpose. For instance, how many seeds does a plant
produce, when just a single one can ensure continuity of its lineage?
It must not be missed that this wastefulness leads to competition and
Darwinian selection.

Before going on to the details of the genetic languages, here is a
quick summary of the components making up the biochemical machinery of
living organisms, at different scales. A framework for understanding
genetic languages must incorporate this hierarchical structure.
\begin{center}
\begin{tabular}{ll}
\hline
Atoms & H,C,N,O, and infrequently P,S \\
Nucleotide bases and amino acids & 10-20 atoms \\
Peptides and drugs & 40-100 atoms \\
Proteins & 100-1000 amino acids \\
Genomes & $10^3$-$10^9$ nucleotide base pairs \\
Size & 1 nm (molecules)-$10^4$ nm (cells) \\
\hline
\end{tabular}
\end{center}
Gene and protein databases have been accumulating a lot of data, which can
be used to test hypotheses and consequences of specific choice of languages.

To summarise, the aim of this article is to understand the physical
and the evolutionary reasons for (a) the specific genetic languages,
and (b) their specific realisations. A tiny footnote is that such
an understanding would have a bearing on the probability of finding
life elsewhere in the universe and then characterising it.

\section{Lessons of Evolution}

Evolution\index{evolution} is the centrepiece of biology. It has been the
cause of many controversies, mainly because it is almost imperceptible---the
evolutionary timescales are orders of magnitude larger than the lifetimes
of individual living organisms. But it is the only scientific principle
that provides a unifying framework encompassing all forms of life, from
the simple origin to an amazing variety. We need to understand the forces
governing the direction of evolution, in order to comprehend where we came
from as well as what the future may have in store for us.

Genetic information\index{genetic information} forms the quantitative
underpinning of evolution. Certain biological facts regarding genetic
languages are well-established:
\begin{enumerate}
\item Languages of genes and proteins are universal.\\
The same 4 nucleotide bases and 20 amino acids are used in DNA, RNA
and proteins, all the way from viruses and bacteria to human beings.
This is despite the fact that other nucleotide bases and amino acids
exist in living cells. This clearly implies that selection of specific
languages has taken place.
\item Genetic information is encoded close to data compression limit
and maximal packing.\\
This indicates that optimisation of information storage has taken place.
\item Evolution occurs through random mutations\index{evolution!mutation},
which are local changes in the genetic sequence. In the long run, however,
only a small fraction of the mutations survive---those proving advantageous
to the organisms.\\
This optimising mechanism is labeled Darwinian
selection\index{evolution!selection}, i.e. competition for limited
resources leading to survival of the fittest.
\end{enumerate}

Over the years, many attempts have been made to construct evolutionary
scenarios that can explain the universality of genetic languages. They
can be broadly classified into two categories. One category is the
``frozen accident\index{frozen accident}" hypothesis \cite{frozen},
i.e. the language somehow came into existence, and became such a vital
part of life's machinery that any change in it would be highly deleterious
to living organisms. This requires the birth of the genetic machinery
to be an extremely rare event, without sufficient time to explore other
possibilities. There is not much room for analysis in this ready-made
solution.  I do not subscribe to it, and instead argue for the other
category. That is the ``optimal solution\index{optimal solution}"
end-point \cite{mathphys}, i.e. the language arrived at its best form
by trial and error, and it did not change thereafter, because any change
in it would make the information processing less competitive. This
requires the evolution of genetic machinery to have sufficient scope
to generate many possibilities, and subsequent competition amongst
them whence the optimal solution wins over the rest.

It should be noted that the existence of an optimising mechanism does
not make a choice between the two categories clear-cut. The reason is
that a multi-parameter optimisation manifold generically has a large
number of minima and maxima, and an optimisation process relying on
only local changes often gets trapped in local minima of the undulating
manifold without reaching the global optimum. In such situations, the
initial conditions and history of evolution become crucial in deciding
the outcome of the process, and typically there arise several isolated
surviving candidates. The globally optimal solution is certainly easier
to reach, when the number of local minima is small and/or the range of
exploratory changes is large. The extent of optimisation is therefore
critically controlled by the ratio of time available for exploration
of various possibilities to the transition time amongst them. For the
genetic machinery to have reached its optimal form, the variety of
possibilities thrown up by the primordial soup must have had a simple
and quick winner.

The procedure of optimisation needs a process of change, and a process
of selection. The former is intrinsic, the latter is extrinsic, and the
two take place at different levels in biology. Indeed the difference
between the two provides much ammunition for debates involving choice
vs. environment, or nature vs. nurture. The changes are provided by
mutations, which occur essentially randomly at the genetic level. That
describes the genotype. The selection takes place by the environmental
pressure at the level of whole organisms. It is not at all random,
rather it is biased towards short-term survival (till reproduction).
That describes the phenotype. We have good reasons to believe that the
primitive living organisms were unicellular, without a nucleus, with
small genomes, and having a simple cellular machinery. In such systems,
the genotype and phenotype levels are quite close, and the early
evolution can easily be considered a direct optimisation problem.

Before exploring what could have happened in the early stages of evolution,
let us also briefly look at the direction\index{evolution!direction} in
which it has continued. The following table summarises how the primitive
unicellular organisms progressed to the level of humans (certainly the
most developed form of life in our own point of view), using different
physical resources to process information at different levels.
\begin{center}
\begin{tabular}{lll}
\hline
Organism      & Messages             & Physical Means  \\
\hline
Single cell   & Molecular            & Chemical bonds, \\
              & (DNA, Proteins)      & Diffusion       \\
\hline
Multicellular & Electrochemical      & Convection,     \\
              & (Nervous system)     & Conduction      \\
\hline
Families,     & Imitation, Teaching, & Light, Sound    \\
Societies     & Languages            &                 \\
\hline
Humans        & Books, Computers,    & Storage devices,\\
              & Telecommunication    & Electromagnetic \\
              &                      & waves           \\
\hline
Gizmos or     & Databases            & Merger of brain \\
Cyborgs ?     &                      & and computer    \\
\hline
\end{tabular}
\end{center}
It is clear that evolution has progressively discovered higher levels of
communication mechanisms, whereby the communication range has expanded
(both in space and time), the physical contact has reduced, abstraction
has increased, succinct language forms have arisen and complex translation
machinery has been developed. More interesting is the manner in which all
this has been achieved, with cooperation\index{evolution!cooperation}
(often with division of labour) gradually replacing competition. This
does not contradict Darwinian selection---it is just that the phenotype
level has moved up, and components of a phenotype are far more likely to
cooperate than compete. The mathematical formulation underlying this
behaviour is ``repeated games", with no foresight but with certain
amount of memory \cite{aumann}.

The evolutionary features useful for the purpose of this article are:\\
$\bullet$ The older and lower information processing levels are far better
optimised than the more recent higher levels. This is a consequence of the
fact that in the optimisation process the lower levels had less options to
deal with and more time to settle on a solution.\\
$\bullet$ The capacity of gathering, using and communicating knowledge
has grown by orders of magnitude in the course of evolution. Indeed one
can surmise that, in the long run, the reach of knowledge overwhelms
physical features in deciding survival fitness.
\begin{quote}
{\it Knowledge\index{knowledge} is the essential driving force behind
evolution,\\ providing a clear direction even when the goal remains unclear.}
\end{quote}

\section{Genetic Languages}

Let us now return to analysing the lowest level of information processing,
i.e. the genetic languages\index{genetic languages}. There are two of
them---the language of DNA and RNA with an alphabet of four nucleotide
bases, and the language of proteins with an alphabet of twenty amino acids.
The tasks carried out by both of them are quite specific and easy to
identify.\\
(1) The essential job of DNA and RNA is to sequentially assemble a chain
of building blocks on top of a pre-existing master template. One can call
DNA the read-only-memory of living organisms. When not involved in the
replication process, the information in DNA remains idle in a secluded
and protected state.\\
(2) Proteins are structurally stable molecules of various shapes and
sizes, with precise locations of active chemical groups. They carry out
various functions of life by highly selective binding to other molecules.
Molecular interactions are weak and extremely short-ranged, and so the
binding necessitates matching of complementary shapes, i.e. lock-and-key
mechanism in three dimensions. Proteins are created whenever needed,
based on the information present in DNA, and disintegrated once their
function is over.

The identification of these tasks makes it easy to see why there are two
languages and not just one. Memory needs long term stability, on the other
hand fast execution of functions is desirable, and the two make different
demands on the hardware involved. (The accuracy of a single language
performing both the tasks would be limited, which is the likely reason
why the RNA world, described later, did not last very long.) Indeed,
our electronic computers compute using electrical signals, but store the
results on the disk using magnetic signals. The former encoding is suitable
for fast processing, while the latter is suitable for long term storage.
The two hardware languages fortunately correspond to the same binary
software language, and are conveniently translated into each other by the
laws of electromagnetism. In case of genetic information, the two hardware
languages work in different dimensions---DNA is a linear chain while
proteins are three dimensional structures---forcing the corresponding
software languages also to be different and the translation machinery
fairly complex.

We want to find the optimal languages for implementing the tasks of
DNA/RNA and proteins. So we have to study what constraints are imposed
on a language for minimisation of errors and minimisation of resources.
Minimisation of errors inevitably leads to a digital language, having a
set of clearly distinguishable building blocks with discrete operations.
With non-overlapping signals, small fluctuations (say less than half the
separation between the discrete values) are interpreted as noise and
eliminated from the message by resetting the values, while large changes
represent genuine change in meaning. The loss of intermediate values is
not a drawback, as long as actual applications need only results with
bounded errors. Minimisation of resources is achieved by using a small
number of building blocks, with simple and quick operations. A versatile
language is then obtained by arranging the building blocks together in
as many different ways as possible.

In this optimisation exercise, the ``minimal language\index{minimal language}",
i.e. the language with the smallest set of building blocks for a given task,
has a unique status \cite{futurecomp}:\\
$\circ$ It has the largest tolerance against errors, since the discrete
variables are spread as far apart as possible in the available range of
physical hardware properties.\\
$\circ$ It has the smallest instruction set, since the number of possible
transformations is automatically limited.\\
$\circ$ It can function with high density of packing and quick operations,
which more than make up for the increased depth of computation.\\
$\circ$ It can avoid the need for translation, by using simple physical
responses of the hardware.

The genetic languages are undoubtedly digital, and that has been crucial
in producing evolution as we know it. Some tell-tale signatures are:\\
$\bullet$ Digital language helps in maintaining variation, while continuous
variables would average out fluctuations.\\
$\bullet$ It is a curious fact that evolution is a consequence of a tiny
error rate. With too many errors the organism will not be able to survive,
but without mutations there will be no evolution.\\
$\bullet$ Even minimal changes in discrete genetic variables generate
sizeable disruptions in the system, and they will be futile unless the
system can tolerate them. Often a large number of trial variations are
needed to find the right combinations, and having only a small number
of discrete possibilities helps. Continuous variables produce gradual
evolution, which appears on larger phenotypic scales when multiple sources
contributing to a particular feature average out.\\
$\bullet$ With most of the trial variations getting rejected as being
unproductive, digital variables give rise to punctuated evolution---sudden
changes interspersed amongst long periods of stasis.

In the following sections, we investigate to what extent the digital genetic
languages are minimal, i.e. we first deduce the minimal languages for the
tasks of DNA/RNA and proteins, and then compare them to what the living
organisms have opted for. A worthwhile bonus is that we gain useful clues
about the simpler predecessors of the modern genetic languages.

\section{Understanding Proteins}

\begin{figure}[b]
\begin{center}
\epsfxsize=10cm
\epsffile{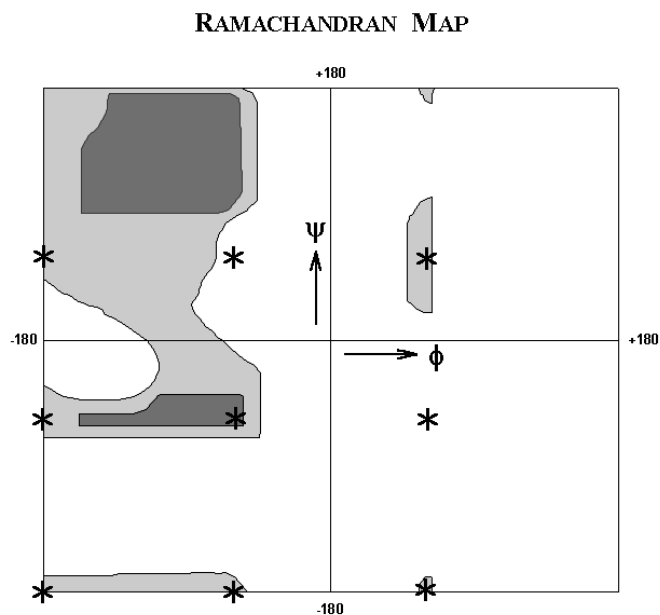}
\end{center}
\vspace{-5mm}
\caption{The allowed orientation angles for the $C_\alpha$ bonds in real
polypeptide chains for chiral L-type amino acids, taking into account
hard core repulsion between atoms [Ramachandran {\it et al.} (1963)].
Stars mark the nine discrete possibilities for the angles, uniformly
separated by $120^\circ$ intervals, when the polypeptide chain is folded
on a diamond lattice.
\label{rmap}}
\end{figure}

Finding the minimal language for proteins\index{minimal language!proteins}
is a straightforward problem in classical geometry \cite{carbon}. The
following is a rapid-fire summary of the analysis.
\begin{itemize}
\item
{\it What is the purpose of the language of amino acids?}\\
To form protein molecules of different shapes and sizes in three
dimensions, and containing different chemical groups.
\item
{\it What is the minimal discrete geometry for designing three
dimensional structures?}\\
Simplicial tetrahedral geometry\index{tetrahedral geometry} and the
diamond lattice. Secondary protein structures, i.e. $\alpha$-helices,
$\beta$-bends and $\beta$-sheets, fit quite well on the diamond lattice.
\item
{\it What are the best physical components to realise this geometry?}\\
Covalently bonded carbon atoms, also $N^+$ and $H_2 O$.
Silicon is far more abundant, but it cannot form aperiodic
structures needed to encode a language.
(In the graphite sheet arrangement, carbon also provides the
simplicial geometry for two dimensional membrane patterns.)
\item
{\it What is a convenient way to assemble these components in the
desired three dimensional structures?}\\
Synthesise one dimensional polypeptide chains, which carry knowledge
about how to fold into three dimensional structures. The problem
then simplifies to assembling one dimensional chains. (Note that
images in our electronic computers are stored as folded sequences.)
\item
{\it What are the elementary operations needed to fold a polypeptide
chain on a diamond lattice, in any desired manner?}\\
Nine discrete rotations, represented as a $3\times3$ array on the
Ramachandran map\index{Ramachandran map} (see Fig.\ref{rmap}).
Additional folding operations are trans-cis flip and long distance bonds.
\item
{\it What can the side groups of polypeptide chains do?}\\
They favour particular orientations of the polypeptide chain by
interactions amongst themselves. They also fill up cavities in the
structure by variations in their size.
\end{itemize}

\begin{figure}[b]
\setlength{\unitlength}{1mm}
\begin{picture}(110,35)
  \thicklines
  \put(21,3){\makebox(0,0)[bl]{(a)}}
  \put(22,19){\makebox(0,0)[bl]{C}}
  \put(21,20){\line(-1, 0){4}}
  \put(10,19){\makebox(0,0)[bl]{H$_2$N}}
  \put(25,20){\line( 1, 0){4}}
  \put(30,19){\makebox(0,0)[bl]{H}}
  \put(23,22){\line( 0, 1){4}}
  \put(22,27){\makebox(0,0)[bl]{COOH}}
  \put(23,18){\line( 0,-1){4}}
  \put(22,11){\makebox(0,0)[bl]{R}}

  \put(70,3){\makebox(0,0)[bl]{(b)}}
  \put(43,19){\circle*{0.5}} \put(44,19){\circle*{0.5}}
  \put(45,19){\circle*{0.5}} \put(46,19){\circle*{0.5}}
  \put(47,18){\makebox(0,0)[bl]{C}}
  \put(48,13){\line( 0,1){4}}
  \put(49,13){\line( 0,1){0.8}}
  \put(49,14.6){\line( 0,1){0.8}} \put(49,16.2){\line( 0,1){0.8}}
  \put(47, 9){\makebox(0,0)[bl]{O}}
  \put(50,20.5){\line( 2,1){4}}
  \put(50,19.5){\circle*{0.5}} \put(51,20){\circle*{0.5}}
  \put(52,20.5){\circle*{0.5}} \put(53,21){\circle*{0.5}}
  \put(54,21.5){\circle*{0.5}}
  \put(55,22){\makebox(0,0)[bl]{N}}
  \put(56.5,26){\line( 0,1){4}}
  \put(55,31){\makebox(0,0)[bl]{H}}
  \put(62,20){\line(-2,1){4}}
  \put(59,20){$\circlearrowright$}
  \put(60,23){$\phi$}
  \put(63,18){\makebox(0,0)[bl]{C$_\alpha$}}
  \put(62,17){\line(-2,-1){4}}
  \put(66,17){\line( 2,-1){4}}
  \put(55,13){\makebox(0,0)[bl]{H}}
  \put(67,15){$\circlearrowright$}
  \put(66,13){$\chi$}
  \put(71,13){\makebox(0,0)[bl]{R}}
  \put(66,20){\line( 2,1){4}}
  \put(67,20){$\circlearrowright$}
  \put(66,23){$\psi$}
  \put(71,22){\makebox(0,0)[bl]{C}}
  \put(72,26){\line( 0,1){4}}
  \put(73,26){\line( 0,1){0.8}}
  \put(73,27.6){\line( 0,1){0.8}} \put(73,29.2){\line( 0,1){0.8}}
  \put(71,31){\makebox(0,0)[bl]{O}}
  \put(78,19.5){\line(-2,1){4}}
  \put(78,20.5){\circle*{0.5}} \put(77,21){\circle*{0.5}}
  \put(76,21.5){\circle*{0.5}} \put(75,22){\circle*{0.5}}
  \put(74,22.5){\circle*{0.5}}
  \put(79,18){\makebox(0,0)[bl]{N}}
  \put(80.5,13){\line( 0,1){4}}
  \put(79, 9){\makebox(0,0)[bl]{H}}
  \put(82,20){\line( 2,1){4}}
  \put(87,22){\makebox(0,0)[bl]{C$_\alpha$}}
  \put(86,25){\line(-2, 1){4}}
  \put(90,25){\line( 2, 1){4}}
  \put(79,27){\makebox(0,0)[bl]{H}}
  \put(95,27){\makebox(0,0)[bl]{R}}
  \put(94,20){\line(-2,1){4}}
  \put(95,18){\makebox(0,0)[bl]{C}}
  \put(96,13){\line( 0,1){4}}
  \put(97,13){\line( 0,1){0.8}}
  \put(97,14.6){\line( 0,1){0.8}} \put(97,16.2){\line( 0,1){0.8}}
  \put(95, 9){\makebox(0,0)[bl]{O}}
  \put(98,20.5){\line( 2,1){4}}
  \put(98,19.5){\circle*{0.5}} \put(99,20){\circle*{0.5}}
  \put(100,20.5){\circle*{0.5}} \put(101,21){\circle*{0.5}}
  \put(102,21.5){\circle*{0.5}}
  \put(103,22){\makebox(0,0)[bl]{N}}
  \put(104.5,26){\line( 0,1){4}}
  \put(103,31){\makebox(0,0)[bl]{H}}
  \put(106,23){\circle*{0.5}} \put(107,23){\circle*{0.5}}
  \put(108,23){\circle*{0.5}} \put(109,23){\circle*{0.5}}
\end{picture}
\vspace{-5mm}
\caption{Chemical structures of (a) amino acid, (b) polypeptide chain.
\label{polypep}}
\end{figure}
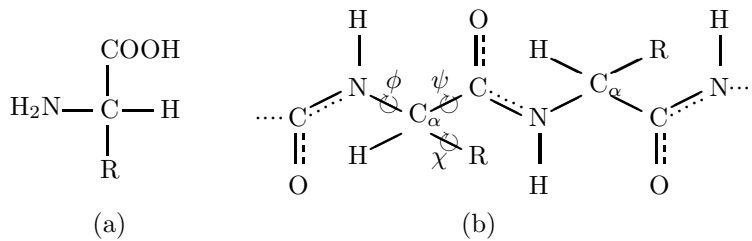

To put the above statements in biological perspective, and to illustrate
the minimalistic choices made by the living organisms (in the context
of what was available), here are some facts about the polypeptide
chains\index{polypeptide chains}.\\
(a) Amino acids\index{amino acids} are easily produced in primordial
chemical soup. They even exist in interstellar clouds.\\
(b) Amino acids are the smallest organic molecules with both an
acid group $(-COOH)$ and a base group $(-NH_2)$. They differ from
each other in terms of distinct R-groups\index{amino acids!R-groups},
which become the side groups of polypeptide chains.\\
(c) Polypeptide chains are produced by polymerisation of amino
acids by acid-base neutralisation (see Fig.\ref{polypep}).\\
(d) Folded $\leftrightarrow$ unfolded transition of polypeptide
chains requires flexible joints and weak non-local interactions
(close to critical behaviour).\\
(e) Transport of polypeptides across membranes is efficient in the
unfolded state than in the folded one, preventing leakage of other
molecules at the same time. (A chain can slide through a small hole.)

The structural language\index{structural language} of the polypeptide
chains would be the most versatile when all possible orientations can
be generated by every amino acid segment. This cannot be achieved by
just a single property of the R-groups (e.g. hydrophobic to hydrophilic
variation). The table below lists the amino acids used by the universal
language of proteins. They are subdivided into several categories
according to the chemical properties of the R-groups, and their molecular
weights provide an indication of the size of the R-groups \cite{lehninger}.
The language of bends and folds of the polypeptide chains is non-local,
i.e. the orientation of an amino acid is not determined by its own R-group
alone, rather the orientation is decided by the interactions of the amino
acid with all its neighbours. Still, by analysing protein databases, one
can find probabilities for every amino acid to participate in specific
secondary structures, and the dominant propensities are listed in the
table as well \cite{creighton}.

\begin{center}
\begin{tabular}{llrrc}
\hline
Amino acid            & R-group   & Mol. wt. & Class & Propensity\\
\hline
G Gly (Glycine)       & Non-polar & 75   & II & turn     \\  %  
A Ala (Alanine)       & aliphatic & 89   & II & $\alpha$ \\  %  
P Pro (Proline)       &           & 115  & II & turn     \\  % d
V Val (Valine)        &           & 117  & I  & $\beta$  \\  % F
L Leu (Leucine)       &           & 131  & I  & $\alpha$ \\  %  
I Ile (Isoleucine)    &           & 131  & I  & $\beta$  \\  % G
\hline
S Ser (Serine)        & Polar     & 105  & II & turn     \\  % f
T Thr (Threonine)     & uncharged & 119  & II & $\beta$  \\  % g
N Asn (Asparagine)    &           & 132  & II & turn     \\  % c
C Cys (Cysteine)      &           & 121  & I  & $\beta$  \\  % D
M Met (Methionine)    &           & 149  & I  & $\alpha$ \\  %  
Q Gln (Glutamine)     &           & 146  & I  & $\alpha$ \\  % C
\hline
D Asp (Aspartate)     & Negative  & 133  & II & turn     \\  % b
E Glu (Glutamate)     & charge    & 147  & I  & $\alpha$ \\  % B
\hline
K Lys (Lysine)        & Positive  & 146  & II & $\alpha$ \\  % a
R Arg (Arginine)      & charge    & 174  & I  & $\alpha$ \\  % A
\hline
H His (Histidine)     & Ring/     & 155  & II & $\alpha$ \\  %  
F Phe (Phenylalanine) & aromatic  & 165  & II & $\beta$  \\  % e
Y Tyr (Tyrosine)      &           & 181  & I  & $\beta$  \\  % E
W Trp (Tryptophan)    &           & 204  & I  & $\beta$  \\  %  
\hline
\end{tabular}
\end{center}

Deciphering the actual orientations of amino acids in proteins is an
outstanding open problem---the protein folding problem\index{protein folding}.
Even then a rough count of the number of amino acids present can be obtained
with one additional input. This is the division of the amino acids into two
classes, according to the properties of the corresponding aminoacyl-tRNA
synthetases (aaRS)\index{aminoacyl-tRNA synthetases}. In the synthesis of
polypeptide chains, tRNA\index{tRNA} molecules are the adaptors with one end
matching with a genetic codon and the other end attached to an amino acid.
The aaRS are the truly bilingual molecules in the translation machinery,
that attach an appropriate amino acid to the tRNA corresponding to its
anticodon. There is a unique aaRS for every amino acid, even though several
different tRNA molecules can carry the same amino acid (the genetic code is
degenerate).  It has been discovered that the aaRS are clearly divided in
two classes, according to their sequence and structural motifs, active sites
and the location where they attach the amino acids to the tRNA molecules
\cite{moras,lewin}. The classes\index{amino acids!classes} of amino acids
are also listed in the table above, and here is what we find:\\
(a) The 20 amino acids are divided into two classes of 10 each.\\
(b) The two classes divide amino acids with each R-group property
equally, in such a way that for every R-group property the larger
R-groups correspond to class I and the smaller ones to class II.\\
(c) The class label of an amino acid can be interpreted as a binary
code for its R-group size, in addition to the categorisation in
terms of chemical properties.\\
(d) This binary code has unambiguous structural significance for
packing of proteins. Folding of an aperiodic chain into a compact
structure invariably leaves behind cavities of different shapes and
sizes. The use of large R-groups to fill big cavities and small
R-groups to fill small ones can produce dense compact structures.\\
(e) Each class contains a special amino acid, involved in operations
other than local folding of polypeptide chains---Cys in class I can
make long distance disulfide bonds, and Pro in class II can induce
trans-cis flip.

We thus arrive at a structural explanation for the 20 amino acids as
building blocks of proteins. Local orientations of the polypeptide
chains have to cover the nine discrete points on the Ramachandran map.
They are governed by the chemical properties of the amino acid R-groups,
and an efficient encoding can do the job with nine amino acids. The
binary code for the R-group sizes fills up the cavities nicely without
disturbing the folds. And then two more non-local operations increase
the stability of protein molecules.

The above counting doesn't tell which sequence of amino acids will lead
to which conformation of the polypeptide chain. That remains an unsolved
exercise in coding as well as chemical properties. On the other hand,
it is known that amino acids located at the active sites and at the
end-points of secondary structures determine the domains and activity of
proteins, while the amino acids in the intervening regions more or less
act like space-fillers. Among the space-fillers, many substitutions
can be carried out that hardly affect the protein function---indeed
protein database analyses have produced probabilistic substitution tables
for the amino acids. We need to somehow incorporate this feature into our
understanding of the structural language of proteins\index{protein folding},
so that we can progress beyond individual letters to words and sentences
[see for example, Socolich {\it et al.} (2005); Russ {\it et al.} (2005)].
A new perspective is necessary, and perhaps the following self-explanatory
paragraph is a clue \cite{rawlinson}. Surprise yourself by reading it at
full speed, even if you are not familiar with crossword puzzles!
\begin{quote}
You arne't ginog to blveiee taht you can aulaclty uesdnatnrd waht I am
wirtnig. Beuacse of the phaonmneal pweor of the hmuan mnid, aoccdrnig to
a rscheearch at Cmabrigde Uinervtisy, it deosn't mttaer in waht oredr
the ltteers in a wrod are, the olny iprmoatnt tihng is taht the frist
and lsat ltteer be in the rghit pclae. The rset can be a taotl mses and
you can sitll raed it wouthit a porbelm. Tihs is bcuseae the huamn mnid
deos not raed ervey lteter by istlef, but the wrod as a wlohe. Amzanig
huh? Yaeh and you awlyas tghuhot slpeling was ipmorantt!
\end{quote}
Written English and proteins are both non-local languages. Evolution,
after all, is no stranger to using a worthwhile idea---here a certain
amount of parallel and distributed processing---over and over again.

\section{Understanding DNA}

Now let us move on to finding the minimal language for DNA and
RNA\index{minimal language!DNA}. Once again, here is a quick-fire
summary of the analysis \cite{quant_gc}.

\begin{itemize}
\item
{\it What is the information processing task carried out by DNA?}\\
Sequential assembly of a complementary copy on top of the pre-existing
template by picking up single nucleotide bases from an unsorted ensemble.
The same task is carried out by mRNA in the assembly of polypeptide chains,
but proceeding in steps of three nucleotide bases (triplet codons).
\item
{\it What is the optimal way of carrying out this task?}\\
Lov Grover's database search algorithm\index{Grover's algorithm}
\cite{grover}, which uses binary queries and requires wave dynamics.
It optimises the number of queries, providing a quadratic speed up over
any Boolean algorithm, irrespective of the size of spatial resources
the Boolean algorithm may use. In a classical wave implementation the
database is encoded as $N$ distinct wave modes, while in a quantum
setting the database is labeled by $\log_2 N$ qubits.
\item
{\it What is the characteristic signature of this algorithm?}\\
The number of queries $Q$ required to pick the desired object from an
unsorted database of size $N$ are given by:
\begin{equation}
\label{querysoln}
(2Q+1) \sin^{-1}\frac{1}{\sqrt{N}} = \frac{\pi}{2}
~\Longrightarrow~\begin{cases}
                         Q=1, &N=4 \cr
                         Q=2, &N=10.5 \cr
                         Q=3, &N=20.2 \cr
                 \end{cases}
\end{equation}
(Non-integral values of $N$ imply small errors in object identification,
about 1 part in 700 and 1050 for $Q=2$ and $3$ respectively.)
\item
{\it What are the physical ingredients needed to implement this algorithm?}\\
A system of coupled wave modes whose superposition maintains phase coherence,
and two reflection operations (phase changes of $\pi$).
\end{itemize}

Again to clarify the biological perspective, and to illustrate the
minimalistic choices made by the living organisms, here are some facts
about the biochemical assembly\index{biochemical assembly} process.\\
(a) Instead of waiting for a desired complex biomolecule to come along,
it is far more efficient to synthesise it from common, simple ingredients.\\
(b) There should be a sufficient number of clearly distinguishable
building blocks to create the wide variety of required biomolecules.\\
(c) The building blocks are randomly floating around in the cellular
environment. They get picked one by one and added to a linearly growing
polymer chain.\\
(d) Complementary nucleotide base-pairing decides the correct building
block to be added at each step of the assembly process.\\
(e) The base-pairings are binary questions; either they form or they do
not form. The molecular bonds involved are hydrogen bonds.

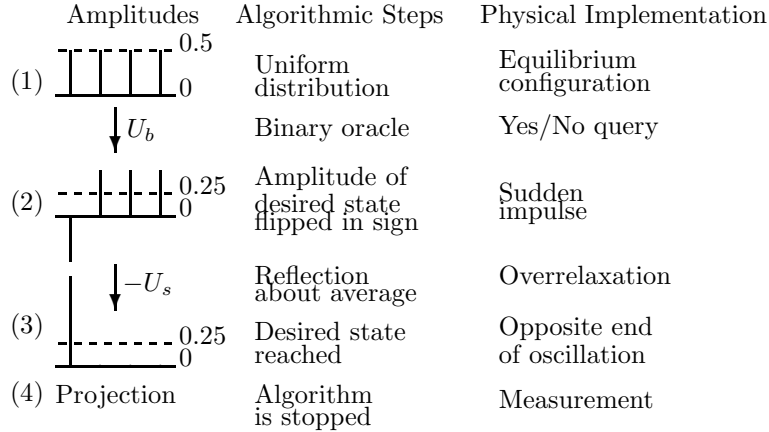
\begin{figure}[t]
\setlength{\unitlength}{0.5mm}
\hspace{5mm}
\begin{picture}(150,120)
  \thicklines
\put(15,105){\makebox(0,0)[bl]{Amplitudes}}
\put(60,105){\makebox(0,0)[bl]{Algorithmic Steps}}
\put(125,105){\makebox(0,0)[bl]{Physical Implementation}}
  \put( 0,87){\makebox(0,0)[bl]{(1)}}
  \put(12,87){\line(1,0){32}}
\put(13,99){\line(1,0){2}} \put(17,99){\line(1,0){2}} \put(21,99){\line(1,0){2}}
\put(25,99){\line(1,0){2}} \put(29,99){\line(1,0){2}} \put(33,99){\line(1,0){2}}
\put(37,99){\line(1,0){2}} \put(41,99){\line(1,0){2}}
  \put(45,87){\makebox(0,0)[bl]{0}} \put(45,99){\makebox(0,0)[bl]{0.5}}
  \put(16,87){\line(0,1){12}} \put(24,87){\line(0,1){12}}
  \put(32,87){\line(0,1){12}} \put(40,87){\line(0,1){12}}
  \put(65,93){\makebox(0,0)[bl]{Uniform}}
  \put(65,87){\makebox(0,0)[bl]{distribution}}
  \put(130,93){\makebox(0,0)[bl]{Equilibrium}}
  \put(130,87){\makebox(0,0)[bl]{configuration}}
  \put(28,84){\vector(0,-1){12}}
  \put(31,75){\makebox(0,0)[bl]{$U_b$}}
  \put(65,75){\makebox(0,0)[bl]{Binary oracle}}
  \put(130,75){\makebox(0,0)[bl]{Yes/No query}}
  \put( 0,55){\makebox(0,0)[bl]{(2)}}
  \put(12,55){\line(1,0){32}}
\put(13,61){\line(1,0){2}} \put(17,61){\line(1,0){2}} \put(21,61){\line(1,0){2}}
\put(25,61){\line(1,0){2}} \put(29,61){\line(1,0){2}} \put(33,61){\line(1,0){2}}
\put(37,61){\line(1,0){2}} \put(41,61){\line(1,0){2}}
  \put(45,55){\makebox(0,0)[bl]{0}} \put(45,61){\makebox(0,0)[bl]{0.25}}
  \put(16,55){\line(0,-1){12}} \put(24,55){\line(0,1){12}}
  \put(32,55){\line(0,1){12}} \put(40,55){\line(0,1){12}}
  \put(65,62){\makebox(0,0)[bl]{Amplitude of}}
  \put(65,56){\makebox(0,0)[bl]{desired state}}
  \put(65,50){\makebox(0,0)[bl]{flipped in sign}}
  \put(130,59){\makebox(0,0)[bl]{Sudden}}
  \put(130,53){\makebox(0,0)[bl]{impulse}}
  \put(28,42){\vector(0,-1){12}}
  \put(30,34){\makebox(0,0)[bl]{$-U_s$}}
  \put(65,37){\makebox(0,0)[bl]{Reflection}}
  \put(65,31){\makebox(0,0)[bl]{about average}}
  \put(130,37){\makebox(0,0)[bl]{Overrelaxation}}
  \put( 0,22){\makebox(0,0)[bl]{(3)}}
  \put(12,15){\line(1,0){32}}
\put(13,21){\line(1,0){2}} \put(17,21){\line(1,0){2}} \put(21,21){\line(1,0){2}}
\put(25,21){\line(1,0){2}} \put(29,21){\line(1,0){2}} \put(33,21){\line(1,0){2}}
\put(37,21){\line(1,0){2}} \put(41,21){\line(1,0){2}}
  \put(45,15){\makebox(0,0)[bl]{0}} \put(45,21){\makebox(0,0)[bl]{0.25}}
  \put(16,15){\line(0,1){24}}
  \put(24,15){\circle*{1}} \put(32,15){\circle*{1}} \put(40,15){\circle*{1}}
  \put(65,22){\makebox(0,0)[bl]{Desired state}}
  \put(65,16){\makebox(0,0)[bl]{reached}}
  \put(130,22){\makebox(0,0)[bl]{Opposite end}}
  \put(130,16){\makebox(0,0)[bl]{of oscillation}}
  \put( 0,4){\makebox(0,0)[bl]{(4)}}
  \put(12,4){\makebox(0,0)[bl]{Projection}}
  \put(65,4){\makebox(0,0)[bl]{Algorithm}}
  \put(65,-2){\makebox(0,0)[bl]{is stopped}}
  \put(130,4){\makebox(0,0)[bl]{Measurement}}
\end{picture}
\caption{The steps of Grover's database search algorithm for the simplest
case of four items, when the first item is desired by the oracle. The left
column depicts the amplitudes of the four states, with the dashed lines
showing their average values. The middle column describes the algorithmic
steps, and the right column mentions their physical implementation.
\label{onefromfour}}
\end{figure}

With these features, the optimal classical algorithm based on Boolean
logic would be a binary tree search. But the observed numbers do not fit
that pattern (of powers of two). On the other hand, the optimal search
solutions of Grover's algorithm are clearly different from and superior
to the Boolean ones, and they do produce the right numbers. The crucial
difference between the two is that wave mechanics\index{wave mechanics}
works with amplitudes and not probabilities, which allows constructive
as well as destructive interference. Grover's algorithm manages the
interference of amplitudes cleverly, and the individual steps are depicted
in Fig.\ref{onefromfour} for the simplest case of four items in the database.

Now note that classically the binary alphabet is the minimal one for
encoding information in a linear chain, and two nucleotide bases (one
complementary pair) are sufficient to encode the genetic information.
As a matter of fact, our digital computers encode all types of information
using only 0's and 1's. The binary alphabet is the simplest system, and so
would have preceded (during evolution) the four nucleotide base system
found in nature. Then, was the speed-up provided by the wave algorithm
the real incentive for nature to complicate the genetic alphabet?
Certainly, if we have to design the optimal system for linear assembly,
knowing all the physical laws that we do, we would opt for something
like what is present in nature. But what did nature really do? We have
no choice but to face the following questions:
\begin{itemize}
\item
{\it Does the genetic machinery have the ingredients to implement
Grover's algorithm\index{Grover's algorithm}?}\\
The physical components are definitely present, and it is not too difficult
to construct scenarios based on quantum dynamics \cite{quant_gc} as well as
vibrational motion \cite{wavesearch}. Although Grover's algorithm was
discovered in the context of quantum computation, it is much more general,
and does not need all the properties of quantum dynamics. In particular,
highly fragile entanglement is unnecessary, while much more stable
superposition\index{superposition} of states is a must. The issue of
concern then is whether coherent superposition of wave modes can survive
long enough for the algorithm to execute. This superposition may be quantum
(i.e. for the wavefunction) or may be classical (as in case of vibrations).
It need not be exactly synchronous either---if the system transits through
all the possible states at a rate much faster than the time scale of the
selection oracle, that would simulate superposition, averaging out high
frequency components (e.g. the appearance of spokes of a rapidly spinning
wheel). Provided that the superposition is achieved somehow, the mathematical
signature, i.e. Eq.(\ref{querysoln}), follows. Explicit formulation of a
testable scenario, based on physical properties of the available molecules
and capable of avoiding fast decoherence\index{decoherence}, is an open
challenge.
\item
{\it Did nature actually exploit Grover's algorithm when the genetic
machinery evolved billions of years ago?}\\
Unfortunately there is no direct answer, since evolution of life cannot
be repeated.
\item
{\it Do the living organisms use Grover's algorithm even today?}\\
In principle, this is experimentally testable. Our technology is yet
to reach a stage where we can directly observe molecular dynamics in
a liquid environment. But indirect tests of optimality are plausible,
e.g. constructing artificial genetic texts containing a different
number of letters and letting it compete with the supposedly optimal
natural language \cite{testdna}.
\end{itemize}
This is not the end of the road, and I return to a deeper analysis later
on. But prior to that let us look at what the above described understanding
of the languages of proteins and DNA has to say about the translation
mechanism between the two, i.e. the genetic code\index{genetic code}.
That investigation does offer non-trivial rewards, regarding how the
complex genetic machinery could have arisen from simpler predecessors.

\section{What Preceded the Optimal Languages?}

Languages of twenty amino acids and four nucleotide bases are too complex
to be established in one go, and evolution must have arrived at them from
simpler predecessors. On the other hand, continuity of knowledge has to be
maintained in evolution from simpler to complex languages, because sudden
drastic changes lead to misinterpretations that kill living organisms.
Two evolutionary routes obeying this restriction, and still capable of
producing large jumps, are known:\\
(1) Duplication of information, which allows one copy to carry on the
required function while the other is free to mutate and give rise to a
new function.\\
(2) Wholesale import of fully functional components by a living organism,
distinct from their own and developed by a different living organism.\\
In what follows, we study the genetic languages within this framework.

\begin{figure}[b]
\begin{center}
\epsfxsize=5cm
\epsffile{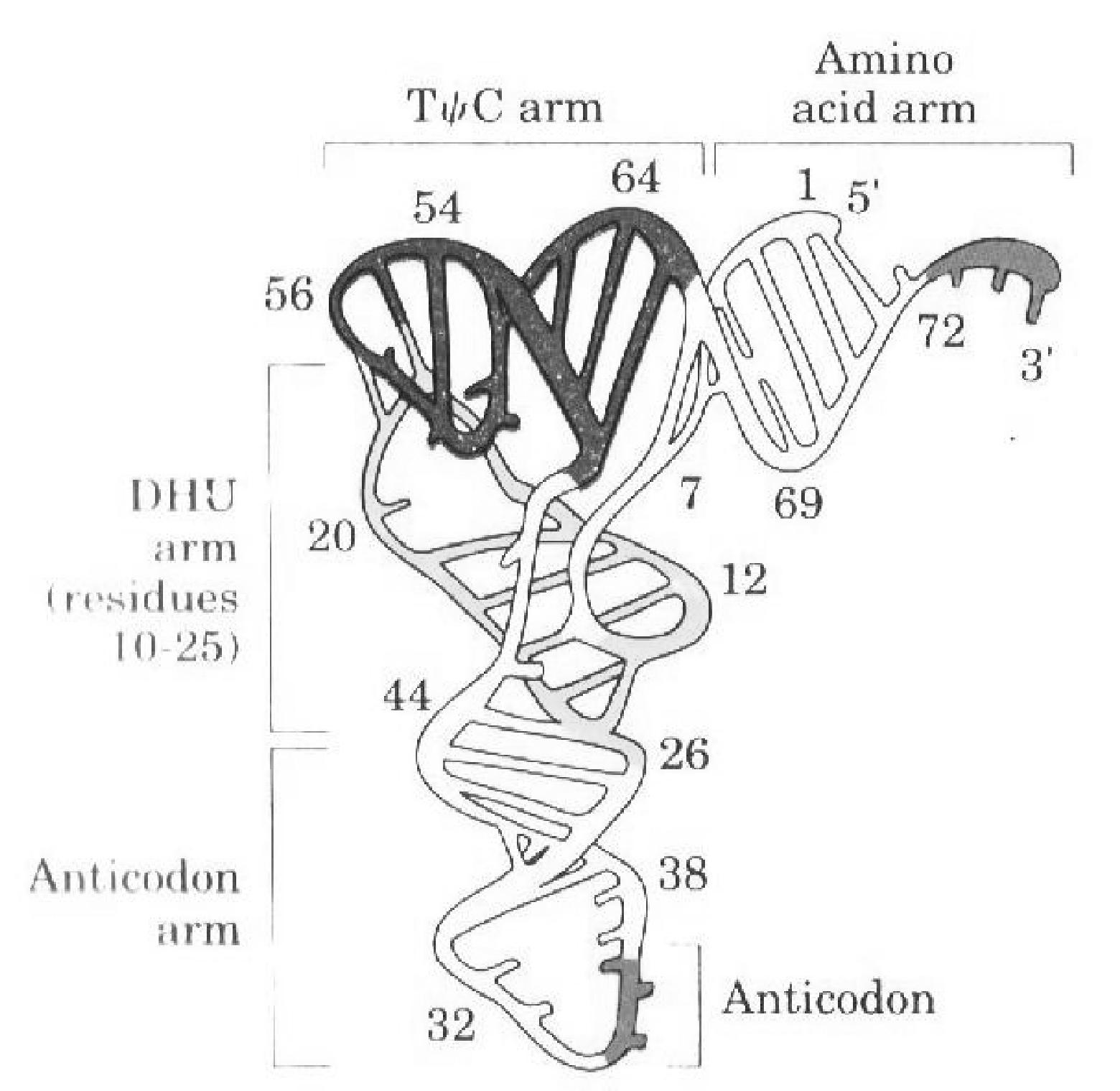}
\hspace{2mm}
\epsfxsize=6cm
\epsffile{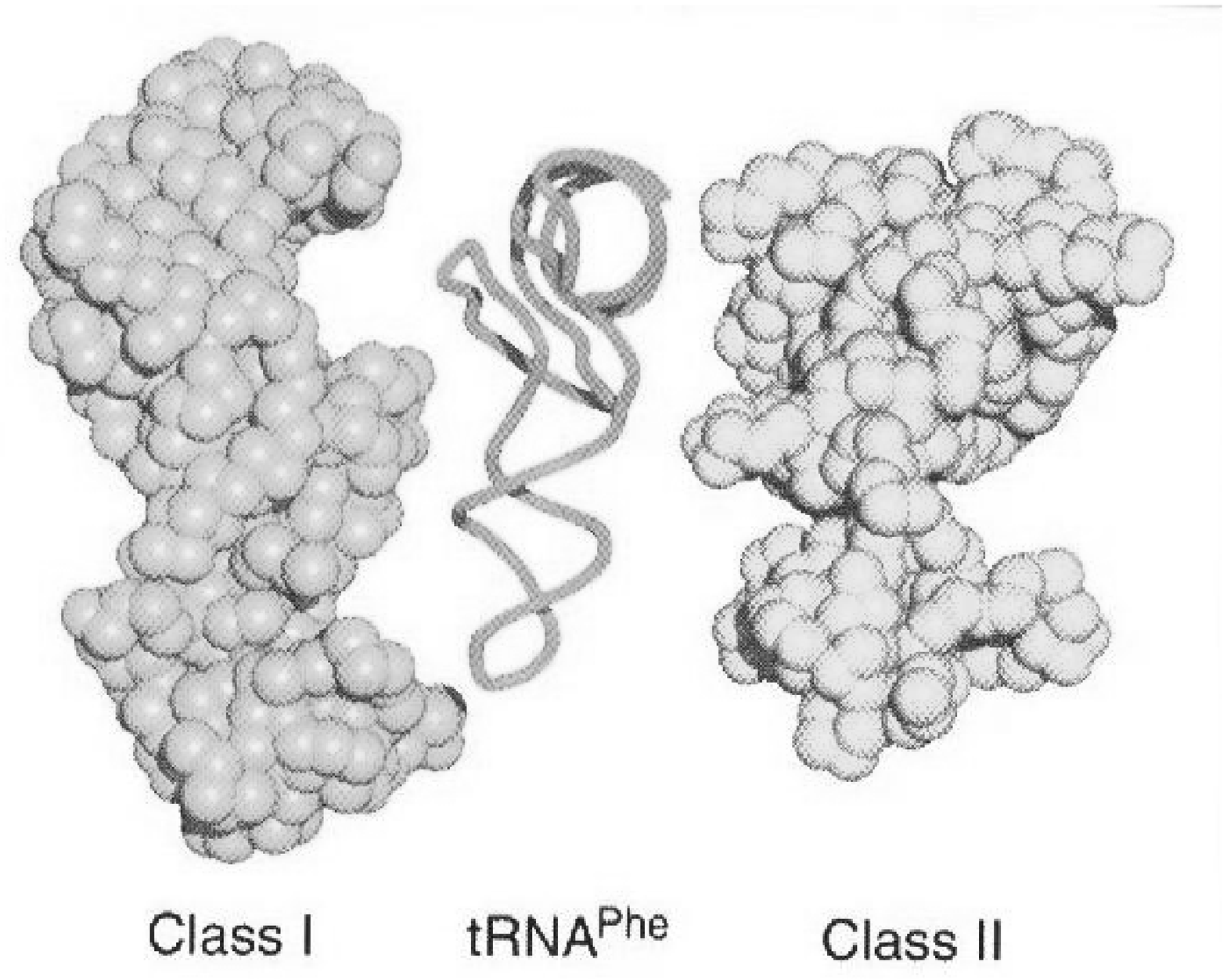}
\end{center}
\caption{The structure of tRNA [Lehninger {\it et al.} (1993)] (left),
and the tRNA-AARS interaction from opposite sides for the two classes
[Arnez and Moras (1997)] (right).
\label{trna}}
\end{figure}

The two classes of amino acids and the $Q=2$ solution of Grover's algorithm,
described in preceding sections, suggest a duplication event\index{genetic
code!duplication}, i.e. the universal non-overlapping triplet genetic code
arose from a more primitive doublet genetic code labeling ten amino acids
\cite{doublet,elsen,rodin}. To justify this hypothesis, we have to identify
evolutionary remnants of
(a) a genetic language where only two nucleotide bases of a codon carry
information while the third one is a punctuation mark,
(b) a set of amino acids that can produce all the orientations of
polypeptide chains but without efficiently filling up the cavities,
and (c) a reasonable association between these codons and amino acids.
Amazingly, biochemical signals for all of these features have been observed.

The central players in this event are the tRNA\index{tRNA} molecules.
They are older than the DNA and the proteins in evolutionary history,
and are believed to link the modern genetic machinery with the earlier RNA
world\index{RNA world} \cite{rnaworld}. It has been discovered that RNA
polymers called ribozymes\index{ribozyme} can both store information and
function as catalytic enzymes, although not very accurately. The hypothesis
is that when more accurate DNA and proteins took over these tasks from
ribozymes, tRNA molecules survived as adaptors from the preceding era.

As illustrated in Fig.\ref{trna}, the tRNAs are L-shaped molecules with
the amino acid acceptor arm at one end and the anticodon arm at the other.
The two arms are separated by a distance of about 75\AA, too far apart for
any direct interaction. The aaRS molecules are much larger than the tRNAs,
and they attach an amino acid to the acceptor stem corresponding to the
anticodon by interacting with both the arms. The two classes\index{amino
acids!classes} of aaRS perform this attachment from opposite sides, in a
mirror image fashion as shown in Fig.\ref{trna}. Class I attachment is from
the minor groove side of the acceptor arm helix, and class II attachment is
from the major groove side\index{amino acids!classes}. It has been observed
that the tRNA acceptor stem sequence, which directly interacts with the
R-group of the amino acid being attached, plays a dominant role in the
amino acid recognition and the anticodon does not matter much. This
behaviour characterises the operational RNA code\index{operational RNA code},
formed by the first four base pairs and the unpaired base N$^{73}$ of the
acceptor stem \cite{schimmel}. The operational code relies on stereochemical
atomic recognition between amino acid R-groups and nucleotide bases; it is
argued to be older than the genetic code and a key to understanding the
goings on in the RNA world.

\begin{center}
The universal genetic code\\
\begin{tabular}{|l|l|l|l|}
\hline
{\bf UUU Phe} & {\bf UCU Ser} &      UAU Tyr  &      UGU Cys  \\
{\bf UUC Phe} & {\bf UCC Ser} &      UAC Tyr  &      UGC Cys  \\
     UUA Leu  & {\bf UCA Ser} &      UAA Stop &      UGA Stop \\
     UUG Leu  & {\bf UCG Ser} &      UAG Stop &      UGG Trp  \\
\hline
     CUU Leu  & {\bf CCU Pro} & {\bf CAU His} &      CGU Arg  \\
     CUC Leu  & {\bf CCC Pro} & {\bf CAC His} &      CGC Arg  \\
     CUA Leu  & {\bf CCA Pro} &      CAA Gln  &      CGA Arg  \\
     CUG Leu  & {\bf CCG Pro} &      CAG Gln  &      CGG Arg  \\
\hline
     AUU Ile  & {\bf ACU Thr} & {\bf AAU Asn} & {\bf AGU Ser} \\
     AUC Ile  & {\bf ACC Thr} & {\bf AAC Asn} & {\bf AGC Ser} \\
     AUA Ile  & {\bf ACA Thr} & {\bf AAA Lys} &      AGA Arg  \\
     AUG Met  & {\bf ACG Thr} & {\bf AAG Lys} &      AGG Arg  \\
\hline
     GUU Val  & {\bf GCU Ala} & {\bf GAU Asp} & {\bf GGU Gly} \\
     GUC Val  & {\bf GCC Ala} & {\bf GAC Asp} & {\bf GGC Gly} \\
     GUA Val  & {\bf GCA Ala} &      GAA Glu  & {\bf GGA Gly} \\
     GUG Val  & {\bf GCG Ala} &      GAG Glu  & {\bf GGG Gly} \\
\hline
\end{tabular}\\
Boldface letters indicate class II amino acids.
\end{center}

We now look at the amino acid class pattern in the genetic code.
The universal triplet genetic code\index{genetic code!universal} has
considerable and non-uniform degeneracy, with 64 codons carrying 21
signals (including Stop) as shown. Although there is a rough rule of
similar codons for similar amino acids, no clear pattern is obvious.

By analysing genomes of living organisms, it has been found that during
the translation process 61 mRNA codons (excluding Stop) pair with a smaller
number of tRNA anticodons. The smaller degeneracy of the anticodons is due
to wobble pairing\index{wobble pairing} of nucleotide bases, where the third
base carries only a limited meaning (either binary or none) instead of
four-fold possibilities \cite{wobble}. The wobble rules are exact for the
mitochondrial code\index{genetic code!mitochondrial}---all that matters
is whether the third base is a purine or a pyrimidine, and the number of
possibilities reduces to 32 as shown. (Note that the mitochondrial code
works with rather small genomes and evolves faster than the universal code,
and so is likely to have simpler optimisation criteria.)

\begin{center}
The (vertebrate) mitochondrial genetic code\\
\begin{tabular}{|l|l|l|l|}
\hline
{\bf UUY Phe} & {\bf UCY Ser} &      UAY Tyr  &      UGY Cys  \\
     UUR Leu  & {\bf UCR Ser} &      UAR Stop &      UGR Trp  \\
\hline
     CUY Leu  & {\bf CCY Pro} & {\bf CAY His} &      CGY Arg  \\
     CUR Leu  & {\bf CCR Pro} &      CAR Gln  &      CGR Arg  \\
\hline
     AUY Ile  & {\bf ACY Thr} & {\bf AAY Asn} & {\bf AGY Ser} \\
     AUR Met  & {\bf ACR Thr} & {\bf AAR Lys} &      AGR Stop \\
\hline
     GUY Val  & {\bf GCY Ala} & {\bf GAY Asp} & {\bf GGY Gly} \\
     GUR Val  & {\bf GCR Ala} &      GAR Glu  & {\bf GGR Gly} \\
\hline
\end{tabular}\\
Boldface letters indicate class II amino acids.\\
Pyrimidines Y=U or C, Purines R=A or G.
\end{center}

The departures exhibited by the mitochondrial genetic code, as well as
the genetic codes of some living organisms, from the universal genetic
code are rather minor, and only occur in some of the positions occupied
by class I amino acids. It can be seen that all the class II amino acids,
except Lys, can be coded by codons NNY and anticodons $\overleftarrow{\rm
GNN}$ (wobble rules allow pairing of G with both U and C) \cite{doublet}.
This pattern suggests that the structurally more complex class I amino
acids entered the genetic machinery later, and a doublet code\index{doublet
code} for the class II amino acids (with the third base acting only as a
punctuation mark) preceded the universal genetic code.

The class pattern becomes especially clear with two more inputs:\\
(1) According to the sequence and structural motifs of their aaRS,
Phe is assigned to class I and Tyr to class II. But if one looks at the
stereochemistry of how the aaRS attach the amino acid to tRNA, then Phe
belongs to class II and Tyr to class I \cite{phe_aars,tyr_aars}. Thus
from the operational RNA code point of view the two need to be swapped.\\
(2) Lys has two distinct aaRS, one belonging to class I (in most
archaea) and the other belonging to class II (in most bacteria and
all eukaryotes) \cite{woese}. On the other hand, the assignment of
AGR codons varies from Arg to Stop, Ser and Gly. This feature is
indicative of an exchange of class roles between AAR and AGR codons
(models swapping Lys and Arg through ornithine have been proposed).\\
These two swaps of class labels do not alter the earlier observation
that the two amino acid classes divide each R-group property equally.
We thus arrive at the predecessor genetic code\index{genetic code!predecessor}
shown below. The binary division of the codons according to the class label
is now not only unmistakable but produces a perfect complementary pattern
\cite{rodin}.

\begin{center}
The predecessor genetic code\\
\begin{tabular}{|l|l|l|l|}
\hline
UUY Phe & {\bf UCY Ser} & {\bf UAY Tyr} &      UGY Cys  \\
UUR Leu & {\bf UCR Ser} &      UAR Stop &      UGR Trp  \\
\hline
CUY Leu & {\bf CCY Pro} & {\bf CAY His} &      CGY Arg  \\
CUR Leu & {\bf CCR Pro} &      CAR Gln  &      CGR Arg  \\
\hline
AUY Ile & {\bf ACY Thr} & {\bf AAY Asn} & {\bf AGY Ser} \\
AUR Met & {\bf ACR Thr} &      AAR Lys* & {\bf AGR Arg*}\\
\hline
GUY Val & {\bf GCY Ala} & {\bf GAY Asp} & {\bf GGY Gly} \\
GUR Val & {\bf GCR Ala} &      GAR Glu  & {\bf GGR Gly} \\
\hline
\end{tabular}\\
Boldface letters indicate class II amino acids.\\
Pyrimidines Y=U or C, Purines R=A or G.
\end{center}
When the middle base is Y (the first two columns), it indicates the
class on its own---U for class I and C for class II. When the middle
base is R (the last two columns), the class is denoted by an additional
Y or R, in the third position when the middle base is A and in the first
position when the middle base is G. (Explicitly the class I codons are
NUN, NAR and YGN, while the class II codons are NCN, NAY and RGN.)
The feature that after the middle base, the first or the third base
determines the amino acid class in a complementary pattern, has led to
the hypothesis that the amino acid class doubling occurred in a strand
symmetric RNA world, with complementary tRNAs providing complementary
anticodons. \cite{rodin}.

The complementary pattern has an echo in the operational code\index{operational
RNA code} of the tRNA acceptor stem\index{tRNA!acceptor stem} too. When the
aaRS attach the amino acid to the $-$CCA tail of the tRNA acceptor arm,
the tail bends back scorpion-like, and the R-group of the amino acid gets
sandwiched between the tRNA acceptor stem groove (bases 1-3 and 70-73) and
the aaRS. Analysis of tRNA consensus sequences from many living organisms
reveals \cite{rodin} that
(a) the first base pair in the acceptor stem groove is almost invariably
G$^1$-C$^{72}$ and is mapped to the wobble position\index{wobble position}
of the codon,
(b) the second base pair is mostly G$^2$-C$^{71}$ or C$^2$-G$^{71}$, which
correlate well respectively with Y and R in the middle position of the codon,
and (c) the other bases do not show any class complementarity pattern.

The involvement of both the operational RNA code and the anticodon in the
selection of appropriate amino acid, and the above mentioned correlations
between the two, make it very likely that the two had a common origin.
Then piecing together all the observed features, the following scenario
emerges for the evolution of the genetic code\index{genetic code!evolution}:\\
(1) Ribozymes\index{ribozyme} of the RNA world could replicate, but their
functional capability was limited---a small alphabet (quite likely four
nucleotide bases) and restricted conformations could only produce certain
types of structures. Polypeptide chains, even with a small repertoire
of amino acids, provided a much more accurate and versatile structural
language, and they took over the functional tasks from ribozymes. This
takeover required close stereochemical matching between ribozymes and
polypeptide chains, in order to retain the functionalities already developed.\\
(2) The class II amino acids provided (or at least dominated) the initial
structural language of proteins. With smaller R-groups, they are easier to
synthesise, and so are likely to have appeared earlier in evolution. They
can fold polypeptide chains in all possible conformations, although some
of the cavities may remain incompletely filled. They also fit snugly into
the major groove of the tRNA acceptor stem\index{tRNA!acceptor stem},
with the bases 1-3 and 70-74 essentially forming a mould for the R-group,
for precise stereochemical recognition. Indeed, this stereochemical
identification of an R-group by three base pairs, necessitated by actual
sizes of molecules, would be the reason for the triplet genetic code,
even in a situation where all the bases do not carry information.\\
(3) The modern tRNA molecules arose from repetitive extensions and
complementary pairing of short acceptor stem sequences. In the process,
the 1-2-3 bases became the forerunners of the 34-35-36 anticodons. With
different structural features identifying the amino acids, paired bases
in the acceptor stem and unpaired bases in the anticodon, the evolution
of the operational code and the genetic code diverged. The two are now
different in exact base sequences, but the purine-pyrimidine label (i.e.
R vs. Y) still shows high degree of correlation between the two.\\
(4) In the earlier era of class II amino acid language, the wobble
base\index{wobble position} was a punctuation mark (likely to be G in
the anticodon, as descendant of the 1-72 pair), the central base was the
dominant identifier (descendant of the 2-71 pair), and the last anticodon
base provided additional specification (equivalent to the 3-70 pair and
the unpaired base 73). During subsequent evolution, these
$\overleftarrow{\rm GNN}$ anticodons have retained their meaning, and all
minor variations observed between genetic codes are in the other anticodons
corresponding to class I amino acids.\\
(5) Class I amino acids got drafted into the structural language, because
they could increase stability of proteins by improved packing of large
cavities without disrupting established structures. The required binary
label for the R-group size, appeared differently in the operational code
and the genetic code. For the operational code, the minor groove of the
acceptor stem was used, and utilisation of the same paired bases from the
opposite side led to a complementary pattern. The class I amino acids fit
loosely in the minor groove, and subsequent proof-reading is necessary at
times to remove incorrectly attached amino acids. For the genetic code,
several of the unassigned anticodons were used for the class I amino acids,
introducing a binary meaning to the wobble position whenever needed. The
Darwinian selection constraint that the operational code and the genetic
code serve a common purpose ensured a rough complementary strand symmetry
for the anticodons as well.\\
(6) The structural language reached its optimal stage, once both classes
of amino acids were incorporated. With 32 anticodons (counting only a
binary meaning for the wobble position) and 20 amino acid signals, enough
anticodons may have remained unassigned. Most of them were taken over by
amino acids with close chemical affinities (wobble position did not assume
any meaning), and a few left over ones mapped to the Stop signal.\\
(7) All this could have happened when each gene was a separate molecule,
coding for a single polypeptide chain. Additional selection pressures
must have arisen when the genes combined into a genome. To take care of
the increased complexity, some juggling of codons happened and the Start
signal appeared. The present analysis is not detailed enough to explain
this later optimisation. Nevertheless, interpretation of similar codons for
similar amino acids and the wobble rules, as relics of the doubling of the
genetic code---indicative but not perfect---is a significant achievement.

At the heart of the class duplication mechanism\index{amino acids!doubling}
described above is (a) the mirror image pattern of the amino acid R-group
fit with the tRNA acceptor stem, and (b) the complementary pattern of the
anticodons. More detailed checks for these are certainly possible. The amino
acids have been tested for direct chemical affinities with either their
codons or their anticodons (but not both together), and most results have
been lukewarm \cite{yarus}. Instead, chemical affinities of amino acids with
paired codon-anticodon grooves should be tested, both by stereochemical models
and actual experiments. It should be also possible to identify which amino
acid paired with which one when the genetic code doubled. Some pairs can be
easily inferred from biochemical properties \cite{ribas,doublet}---(Asp,Glu),
(Asn,Gln), (Lys,Arg), (Pro,Cys), (Phe,Tyr), (Ser\&Thr,Val\&Ile)---while the
others would be revealed by stereochemical modeling.

The next interesting exercise, further back in time and therefore more
speculative, is to identify how a single class 10 amino acid language
took over the functional tasks of 4 nucleotide base ribozymes. This is
the stage where Grover's algorithm might have played a crucial role,
and so we go back and look into it more inquisitively. 

\section{Quantum Role?}

The arguments of the preceding section reduce the amino acid identification
problem by a triplet code, to the identification problem within a class by
a doublet code plus a binary class label. It is an accidental degeneracy
that the $Q=3$ solution of Grover's algorithm\index{Grover's algorithm},
Eq.(\ref{querysoln}), can be obtained as the $Q=2$ solution plus a classical
binary query. To assert that the sequential assembly\index{biochemical
assembly} process reached its optimal solution, we still need to resolve
how the $Q=1,2$ solutions of Eq.(\ref{querysoln}) were realised by the
primordial living organisms.

Clearly, the assembly processes occur at the molecular scale. We know
the physical laws applicable there---classical dynamics is relevant, but
quantum dynamics cannot be bypassed. Discrete atomic structure provided by
quantum mechanics is the basis of digital genetic languages. Molecular bonds
are generally given a classical description, but they cannot take place
without appropriate quantum correlations among the electron wavefunctions. 
Especially, hydrogen bonds are critical to the genetic identification
process, and they are inherently quantum---typical examples of tunneling
in a double well potential. The assemblers, i.e. the polymerase enzymes and
the aaRS molecules, are much larger than the nucleotide bases and the amino
acids, and completely enclose the active regions where identification of
nucleotide bases and amino acids occurs. They provide a well-shielded
environment for the assembly process, but the cover-up also makes it
difficult to figure out what exactly goes on inside.

Chemical reactions are typically described in terms of specific initial
and final states, and transition matrix elements between the two that
characterise the reaction rates. That is a fully classical description,
and it works well for most practical purposes. But to the best of our
understanding, the fundamental laws of physics are quantum and not
classical---the classical behaviour arises from the quantum world as an
``averaged out" description. Quantum steps are thus necessarily present
inside averaged out chemical reaction rates, and would be revealed if we
can locate their characteristic signatures. In the present context, such
a fingerprint is superposition\index{superposition}.

The initial and final states of Grover's algorithm are classical, but
the execution in between is not. In order to be stable, the initial and
final states have to be based on a relaxation towards equilibrium process.
For the execution of the algorithm in between, the minimal physical
requirement is a system that allows superposition of states, in particular
a set of coupled wave modes. As illustrated earlier in Fig.\ref{onefromfour},
the algorithm needs two reflection operations. Provided that the necessary
superposition is achieved somehow, it is straightforward to map these
operations to: (i) the impulse interaction during molecular bond formation
which has the right properties to realise the selection oracle as a fairly
stable geometric phase, and (ii) the (damped) oscillations of the subsequent
relaxation, which when stopped at the right instant by release of the binding
energy to the environment can make up the other reflection phase.

Beyond this generic description, the specific wave modes to be superposed
can come from a variety of physical resources, e.g. quantum evolution,
vibrations and rotations. With properly tuned couplings, resonant transfer
of amplitudes occurs amongst the wave modes (the phenomenon of beats), and
that is the dynamics of Grover's algorithm. When the waves remain coherent,
their amplitudes add and subtract, and we have superposition. But when the
waves lose their coherence, we get an averaged out result---a classical
mixture. Thus the bottom line of the problem is:
\vspace{-2mm}
\begin{quote}
{\it Can the genetic machinery maintain coherence of appropriate wave
modes on a time scale required by the transition matrix elements?}
\end{quote}

Explicitly, let $t_b$ be the time for molecular identification by bond
formation, $t_{coh}$ be the time over which coherent superposition holds,
and $t_{rel}$ be the time scale for relaxation to equilibrium. Then,
Grover's algorithm can be executed when the time scales satisfy the
hierarchy
\begin{equation}
t_b \ll t_{coh} \ll t_{rel} ~.
\end{equation}
Other than this constraint, the algorithm is quite robust and does not
rely on fine-tuned parameters. (Damping is the dominant source of error;
other effects produce errors which are quadratic in perturbation parameters.)

Wave modes inevitably decohere due to their interaction with environment,
essentially through molecular collisions and long range forces.
Decoherence\index{decoherence} always produces a cross-over leading to
irreversible loss of information \cite{zeh}---collapse of the wavefunction
in the quantum case and damped oscillations for classical waves. The time
scales of decoherence depend on the dynamics involved, but a generic
feature is that no wave motion can be damped faster than its natural
undamped frequency of oscillation. For an oscillator,
\begin{equation}
\label{critdamp}
\ddot{x} + 2\gamma\dot{x} + \omega_0^2 x = 0,~ x \sim e^{i\omega t}
~\Longrightarrow~ \gamma_{\rm crit} = {\rm max}({\rm Im}(\omega)) = \omega_0 ~.
\end{equation}
Too much damping freezes the wave amplitude instead of making it decay.
Thus $\omega_0^{-1}$ is both an estimate of $t_b$ and a lower bound on
$t_{coh}$. Molecular properties yield
$\omega_0 = \Delta E/\hbar = O(10^{14}){\rm sec}^{-1}$,
for the transition frequencies of weak bonds as well as for the vibration
frequencies of covalent bonds.

Decoherence must be controlled in order to observe wave dynamics,
irrespective of any other (undiscovered) physical phenomena that may
be involved. In case of vibrational and rotational modes of molecules,
the fact that we can experimentally measure the excitation spectra
implies that the decoherence times are much longer than $t_b$.
In case of quantum dynamics, the decoherence rate is often estimated from
the scattering cross-sections of environmental interactions, in dilute gas
approximation using conventional thermodynamics and Fermi's golden rule.
For molecular processes, these times are usually minuscule, orders of
magnitude below $\omega_0^{-1}$. In view of Eq.(\ref{critdamp}), such
minuscule estimates are wrong---the reason being that Fermi's golden
rule is an approximation, not valid at times smaller than the natural
oscillation period. A more careful analysis is necessary.

According to Fermi's golden rule, the environmental decoherence rate is
inversely proportional to three factors: the initial flux, the interaction
strength and the final density of states. We know specific situations,
where quantum states are long-lived due to suppression of one or more of
these factors. The initial flux is typically reduced by low temperatures
and shielding, the interaction strength is small for lasers and nuclear
spins, and the final density of states is suppressed due to energy gap for
superconductors and hydrogen bonds. We need to investigate whether or not
these features are exploited by the genetic machinery, and if so to what
extent.

Large catalytic enzymes\index{enzymes} (e.g. polymerases, aaRS, ribosomes)
have an indispensable role in biomolecular assembly processes. These
processes do not take place in thermal equilibrium, rather the enzymes
provide an environment that supplies free energy (using ATP molecules)
as well as shields. The assembly then proceeds along the chain linearly
in time. In a free solution without the enzymes, the assembly just does
not take place, even though such a free assembly would have the advantage
of parallel processing (i.e. simultaneous assembly all along the chain).
The enzymes certainly reduce the external disturbances and decrease the
final density of states by limiting possible configurations. But much more
than that, they stabilise the intermediate reaction states, called the
transition states\index{transition states}. The traditional description is
that the free energy barrier between the reactants and the products is too
high to cross with just the thermal fluctuations, and the enzymes take the
process forward by lowering the barrier and supplying free energy. The
transition states are generally depicted using distorted electron clouds,
somewhere in between the configurations of the reactants and the products,
and they are unstable when not assisted by the enzymes. They can only be
interpreted as superpositions, and not as mixtures---we have to accept
that the enzymes stabilise such intermediate superposition states while
driving biomolecular processes.

Thus we arrive at the heart of the inquiry:
\vspace{-2mm}
\begin{quote}
{\it Grover's algorithm needs certain type of superpositions, and catalytic
enzymes can stabilize certain type of superpositions. Do the two match, and
if so, what is the nature of this superposition?}
\end{quote}

The specific details of the answer depend on the dynamical mechanism
involved. The requisite superposition is of molecules that have a largely
common structure while differing from each other by about 5-10 atoms.
I have proposed two possibilities \cite{quant_gc,wavesearch}:\\
(1) In a quantum scenario, wavefunctions get superposed and the algorithm
enhances the probability of finding the desired state. Chemically distinct
molecules cannot be directly superposed, but they can be effectively
superposed by a rapid cut-and-paste job of chemical groups (enzymes are
known to perform such cut-and-paste jobs). Whether this really occurs,
faster than the identification time scale $t_b$ and with the decoherence
time scale significantly longer than $\hbar/\omega_0$, is a question that
should be experimentally addressed. It is a tough proposition, and most
theoretical estimates are pessimistic.\\
(2) In a classical wave scenario, all the candidate molecules need to be
present simultaneously and coupled together in a specific manner. The
algorithm concentrates mechanical energy of the system into the desired
molecule by coherent oscillations, helping it cross the energy barrier
and complete the chemical reaction. Enzymes are required to couple the
components together with specific normal modes of oscillation, and long
enough coherence times are achievable. This scenario provides the same
speed up in the number of queries $Q$ as the quantum one, but involves
extra spatial costs. The extra cost is not insurmountable in the small
$N$ solutions relevant to genetic languages, and the extra stability
against decoherence makes the classical wave scenario preferable. (Once
again note that time optimisation is far more important in biology than
space optimisation.)

Twists and turns can be added to these scenarios while constructing a
detailed picture. But in any implementation of Grover's algorithm, the
requirement of superposition would manifest itself as simultaneous presence
of all the candidate molecules during the selection process, in contrast to
the one-by-one trials of a Boolean algorithm. This particular aspect can be
experimentally tested by the available techniques of isotope substitution,
NMR spectroscopy and resonance frequency measurements. The algorithm also
requires the enzymes to play a central role in driving the non-equilibrium
selection process, but direct observation of that would have to await
breakthroughs in technologies at nanometre and femtosecond scales.

\section{Outlook}

Information theory\index{information theory} provides a powerful framework
for extracting essential features of complicated processes of life, and then
analysing them in a systematic manner. The easiest processes to study are no
doubt the ones at the lowest level. We have learned a lot, both in computer
science and in molecular biology, since their early days \cite{schrodinger,
neumann,frozen}, and so we can now perform a much more detailed study.
Physical theories often start out as effective theories, where predictions
of the theories depend on certain parameters. The values of the parameters
have to be either assumed or taken from experiments; the effective theory
cannot predict them. To understand why the parameters have the values they
do, we have to go one level deeper---typically to smaller scales. When the
deeper level reduces the number of unknown parameters, we consider the theory
to be more complete and satisfactory. The level below conventional molecular
biology is spanned by atomic structure and quantum dynamics, and that is the
natural place to look for reasons behind life's ``frozen accident". It is
indeed wonderful that sufficient ingredients exist at this deeper level to
explain the frozen accident as the optimal solution. The first reward of
this analysis has been a glimpse of how the optimal solution was arrived at.

Evolution\index{evolution} of life occurs through random events (i.e.
mutations), without any foresight or precise rules of logic. It is the
powerful criterion of survival, in a usually uncomfortable and at times
hostile environment, that provides evolution a direction. Even though we do
not really understand why living organisms want to perpetuate themselves,
we have enough evidence to show that they use all available means for this
purpose \cite{dawkins}. This struggle for fitness allows us to assign
underlying patterns to evolution---not always perfect, frequently with
variations, and yet very much practical. By understanding these patterns,
we can narrow down the search for a likely evolutionary route among a
multitude of possibilities. Such an insight is invaluable when we want
to extrapolate in the unknown past with scant direct evidence. That is
certainly the case in trying to understand the origin of life as we know
it. Of course, the inferences become stronger when supported by simulated
experiments, and worthwhile tests of every hypothesis presented have been
pointed out in the course of this article.

Counting the number of building blocks in the languages of DNA and proteins,
and finding patterns in them, is only the beginning of a long exercise to
master these languages. Natural criteria for the selection of particular
building blocks would be chemical simplicity (for easy availability and
quick synthesis) and functional ability (for implementing the desired tasks).
Life can be considered to have originated, not with just complex chemical
interactions in a primordial soup, but only when the knowledge of functions
of biomolecules started getting passed from one generation to the next.
This logic puts the RNA world\index{RNA world} before the modern genetic
machinery; ribozymes provide both function and memory, to a limited extent
but with simpler ingredients. During evolution, the structurally more
versatile polypeptides---they have been observed to successfully mimic DNA
\cite{walkinshaw} as well as tRNA \cite{nakamura}---took over the task of
creating complex biochemistry, while leaving the memory storage job to DNA.
The work described in this article definitely reinforces this point of view,
with simpler predecessors of the modern genetic languages to be found in the
stereochemical interaction between the tRNA acceptor stem and simple class II
amino acids. Experimental verification of this hypothesis would by and large
solve the translation mystery, i.e. which amino acid corresponds to which
codon/anticodon. Then we can push the analysis further back in time, to the
still simpler language of ribozymes, and try to figure out what went on in
the RNA world.

The opposite direction of investigation, of constructing words and sentences
from the letters of alphabets, is much more than a theoretical adventure and
closely tied to what the future holds for us. We want to design biomolecules
that carry out specific tasks, and that needs unraveling how the functions
are encoded in the three dimensional protein assembly process. This is a 
tedious and difficult exercise, involving hierarchical structures and
subjective variety. But some clues have appeared, and they should be built
on to understand more and more complicated processes of life. We may feel
uneasy and scared about consequences of redesigning ourselves, but that
after all would also be an inevitable part of evolution!

\vspace{5mm}
% Sing a song of sixpence a pocket full of rye,
% four and twenty blackbirds baked in a pie.
\rightline{\it ``\ldots when the pie was opened, the birds began to sing \ldots"}
% Now, wasn't that a dainty dish to set before the king?

\vfill
\noindent
{\bf About the author}
\begin{quote}
{\bf Apoorva D. Patel}
is a Professor at the Centre for High Energy Physics, Indian Institute
of Science, Bangalore. He obtained his M.Sc. in Physics from the Indian
Institute of Technology, Mumbai (1980), and Ph.D. in Physics from the
California Institute of Technology (Caltech) under Geoffrey Fox (1984).
His major field of work has been the theory of QCD, where he has used
lattice gauge theory techniques to investigate spectral properties,
phase transitions and matrix elements. In recent years, he has worked on
quantum algorithms, and used information theory concepts to understand
the structure of genetic languages.
\end{quote}

\printindex

\end{document}